\RequirePackage{amsthm}

\documentclass[pdflatex,sn-nature]{sn-jnl}

\usepackage{graphicx}%
\usepackage{multirow}%
\usepackage{amsmath,amssymb,amsfonts}%
\usepackage{amsthm}%
\usepackage{mathrsfs}%
\usepackage[title]{appendix}%
\usepackage{xcolor}%
\usepackage{textcomp}%
\usepackage{manyfoot}%
\usepackage{booktabs}%
\usepackage{algorithm}%
\usepackage{algorithmicx}%
\usepackage{algpseudocode}%
\usepackage{listings}%
\usepackage{hyperref}
\usepackage[all]{hypcap} 
\usepackage{anyfontsize}
\usepackage[section]{placeins}
\evensidemargin\oddsidemargin
\theoremstyle{thmstyleone}%
\theoremstyle{thmstyletwo}%

\theoremstyle{thmstylethree}%

\begin{document}

\title[Mapping the discrete folding landscape]{Mapping the discrete folding landscape}


\author*[1,2]{\fnm{João} \sur{C. Neves}}\email{jlneves@fc.ul.pt}

\author*[1,2]{\fnm{Bernardo} \sur{R. Marques}}\email{fc59924@alunos.ciencias.ulisboa.pt}

\author[1,2]{\fnm{Cristóvão} \sur{S. Dias}}\email{csdias@fc.ul.pt}
\equalcont{These authors contributed equally to this work.}

\author[1,2]{\fnm{Nuno} \sur{A. M. Araújo}}\email{nmaraujo@fc.ul.pt}
\equalcont{These authors contributed equally to this work.}

\affil*[1]{\orgdiv{Departamento de Física}, \orgname{Faculdade de Ciências, Universidade de Lisboa}, \orgaddress{\city{Lisboa}, \postcode{1749-016}, \country{Portugal}}}

\affil[2]{\orgdiv{Centro de Física Teórica e Computacional}, \orgname{Universidade de Lisboa}, \orgaddress{\city{Lisboa}, \postcode{1749-016}, \country{Portugal}}}


\abstract{Folding is emerging as a promising manufacturing process to transform flat materials into functional structures, offering efficiency by reducing the need for welding, gluing, and molding, while minimizing waste and enabling automation. Designing target shapes requires not only to determine cuts and folds, but also folding pathways. Simple combinatorics is impractical as the possibilities grow factorially with the number of folds. To address this, we present a graph-based algorithm for polyhedral shapes. By representing the target shape as a graph, where nodes correspond to faces and edges represent adjacency, the algorithm identifies all possible fold sequences and maps the configuration space into a discrete set of intermediate configurations. This systematic mapping is critical for the design of optimized processes, the simplifying of folding operations, the reduction of failures, and the improvement of manufacturing reliability.}




\maketitle

\section{Introduction}\label{sec1}

Folding is becoming increasingly relevant in modern manufacturing due to its remarkable versatility \cite{meta_materials_review1, meta_materials_review2}. By enabling intricate 3D shapes from flat materials while reducing the need for adhesives and joints, folding significantly minimizes material waste compared to conventional methods \cite{Chen2020}. It also offers exceptional flexibility, supports automation, and facilitates rapid prototyping through laser cutting and 3D printing technologies. These benefits make folding a transformative tool in industries such as robotics, packaging, aerospace, and biomedical engineering \cite{Felton2014, Gu2023, Pedivellano2024, Zirbel2013, mitani2024, Zhu2024, Babaee2021, Huang2019, McMullen2022}.

\begin{figure}[!htb]
        \centering
        \includegraphics[width=\textwidth]{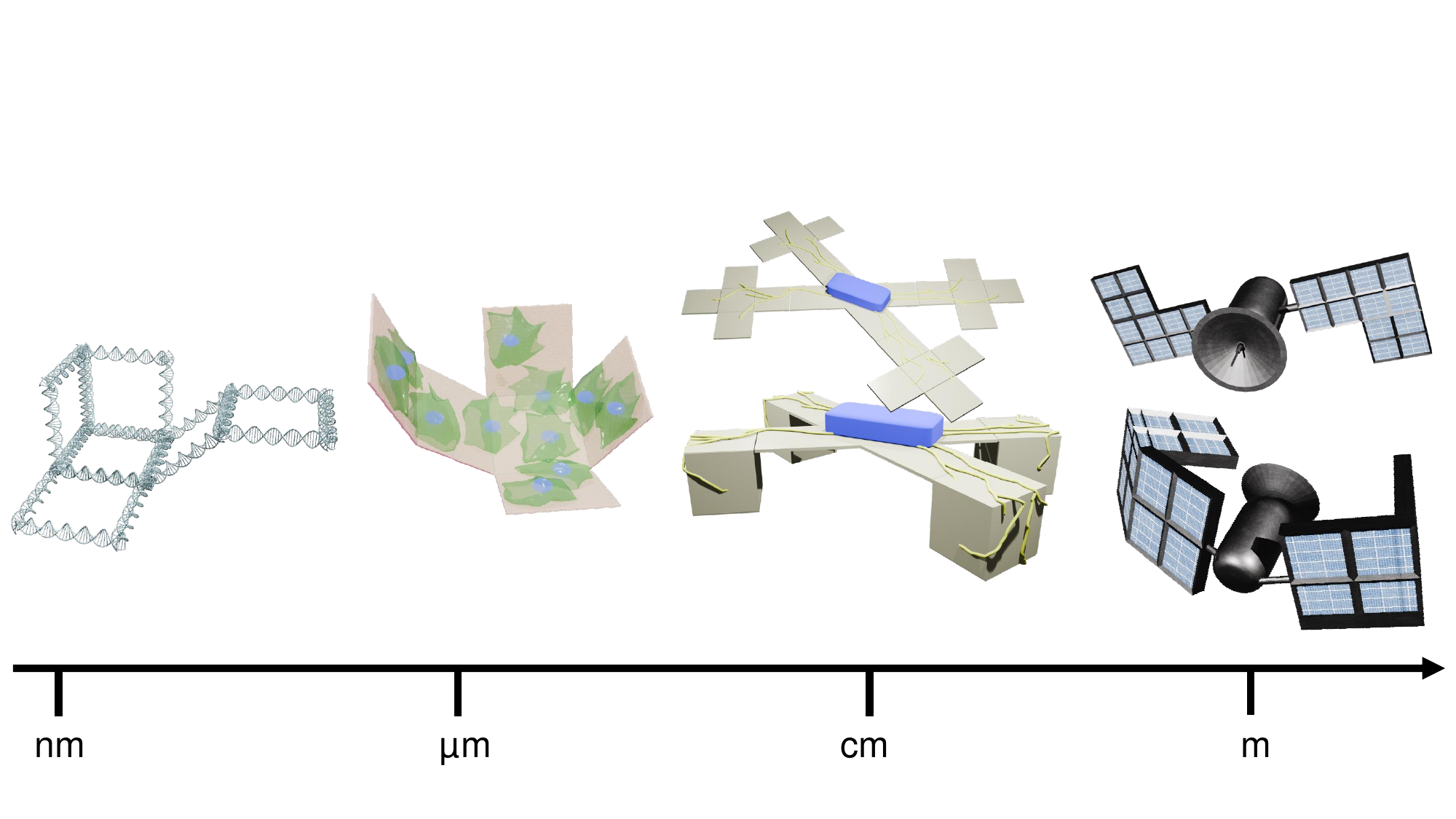}
        \caption{\textbf{Examples of foldable structures at different scales.} From DNA strands origami structures at the nanometer scale \cite{Veneziano2020, Kim2023}, to cell-activated hinges \cite{Cells2012}, to self-folding robots \cite{Felton2014} and to large-scale foldable telescopes.}
        \label{kirigami_apps}
\end{figure}

The idea of using folding to build structures applies across various length scales (see Fig.~\ref{kirigami_apps}), from nanoscale bioengineered DNA networks \cite{Veneziano2020, Kim2023} to macroscale foldable satellites \cite{Felton2014}. Microscale applications include cell-activated hinges \cite{Cells2012}, while centimeter-scale innovations feature self-folding robots \cite{Felton2014}. Additional applications span soft robotics \cite{Felton2014, Gu2023}, biomedical devices \cite{Gu2023, Babaee2021, Huang2019}, space technology \cite{Pedivellano2024, Zirbel2013}, and drug delivery systems \cite{Cells2012, Lamoureux2015}. Folding methods have also been used to produce meta-materials using strategies such as creating a well-defined pattern of folds like in the Miura-Ori \cite{Yasuda2015, Wang2017, Fang2018}, cuts \cite{Tao2023} or by creating foldable sub-structures which are modular and can be combined to produce more complex shapes \cite{Li2021}. Foldable meta-materials provide functions ranging from ultra-stiffness \cite{Zheng2014} and negative Poisson ratios \cite{Yasuda2015} to negative thermal expansion \cite{Guo2021}, impact absorption \cite{Li2020, Zhang2022, Fathers2015}, and the development of programmable \cite{Sussman2015, An2020} or tunable shapes \cite{Neville2016}. While folding mechanisms and binding strategies vary across scales and systems, some features, such as the set of intermediate configurations, depend solely on geometry and topology, making them independent of scale or material.

The concept of folding 3D shapes from flat (2D) templates originates from the ancient art of origami, in which paper (\textit{gami}) is folded into 3D shapes through precise sequences of folds (\textit{ori}). The introduction of cuts (\textit{kiri}), as in kirigami, expands this capability to more intricate shapes. A classic example is the Latin cross template, consisting of a row of four adjacent squares with additional squares on each side of the second square, which forms a cube when folded along its edges (see Fig.~\ref{algorithm_explanation}.A). Such templates, also known as nets, represent the 2D unfolding of 3D structures and were first formalized in the 16th century by Albrecht Dürer \cite{O’Rourke_2011}.

Significant research has focused on identifying templates for specific folded structures and predicting the resulting 3D shapes. For polyhedral shapes, templates can be obtained by edge unfolding, where a set of cuts along edges allows the structure to flatten into a single piece \cite{Demaine_O’Rourke_2007}. However, these cuts must respect certain constraints: they must not form a loop to prevent multiple disconnected pieces, and they must visit all vertices to ensure a flat unfolded structure. Determining intermediate configurations and folding pathways, however, remains challenging. If we assume that all configurations with the same set of connected faces (uncut edges) are equivalent, regardless of the angles between faces along the free-moving edges, then the configuration landscape becomes discrete. Once the required edges for unfolding are known, identifying intermediate structures reduces to evaluating all possible subsets of these cuts. However, the number of possibilities grows factorially with the number of edges, making the task impractical. Even for a cube, which requires cutting along only seven edges, there are 5,040 possible sequences to be tested. For an icosahedron, with just four more cuts, the number of sequences explodes to 39 million. In addition, cutting along an edge produces a new intermediate configuration but does not necessarily lead to a new folding state, as the number of free-moving edges remains the same. For example, in a cube, at least three edges must be cut along before a face can unfold, an aspect that straightforward combinatorial analysis fails to capture. Alternatively, direct simulations of foldings can be attempted, but these methods often fail to account for kinetically improbable structures, leaving many configurations unexplored \cite{Dodd2018}. The current methods for probing the configurational space rely on searching for vertex connections on templates and sequentially connect them, leading to intermediate structures and therefore pathways of folding. This approach, however selects only a subset of intermediate structures which despite being kinetically more probable do not correspond to the whole set of intermediate structures \cite{Kaplan2014, JohnsonChyzhykov2016}. Our method is able to completely map the configurational space given an initial and target structure while maintatining its efficiency much below the factorial explosion (See Supplementary Material \ref{A4_TimeEfficiency}).

It has been shown that both shapes and templates can be mapped onto graphs \cite{Araujo2018, Demaine_O’Rourke_2007, Dodd2018}. One graph is the \textit{face graph}, where the nodes are the faces and links indicate adjacency. The other is the \textit{shell graph}, where the nodes correspond to the vertices of the polyhedron, and links represent its edges. Since different structures yield different graph representations, adding or removing links in a graph corresponds to folding or unfolding faces. Thus, the face graphs of the shape and of the corresponding templates share the same nodes and differ only in the number of links. Within the graph representation framework, templates can be uniquely and deterministically obtained by identifying spanning trees in the \textit{face graph} of the shape \cite{Araujo2018}. To identify intermediate configurations, we first determine the set of cuts that lead to unfolding while accounting for previously visited configurations, reducing the number of evaluations needed. Our algorithm, given any initial and final structure, searches for all possible folding paths and intermediate configurations. This approach allows us to construct a complete catalog of all possible intermediate configurations and to map all paths between them, thereby representing the discrete folding landscape as a directed graph.

\section{Model Description}

\begin{figure}[!htb]
        \centering
        \includegraphics[width=\textwidth]{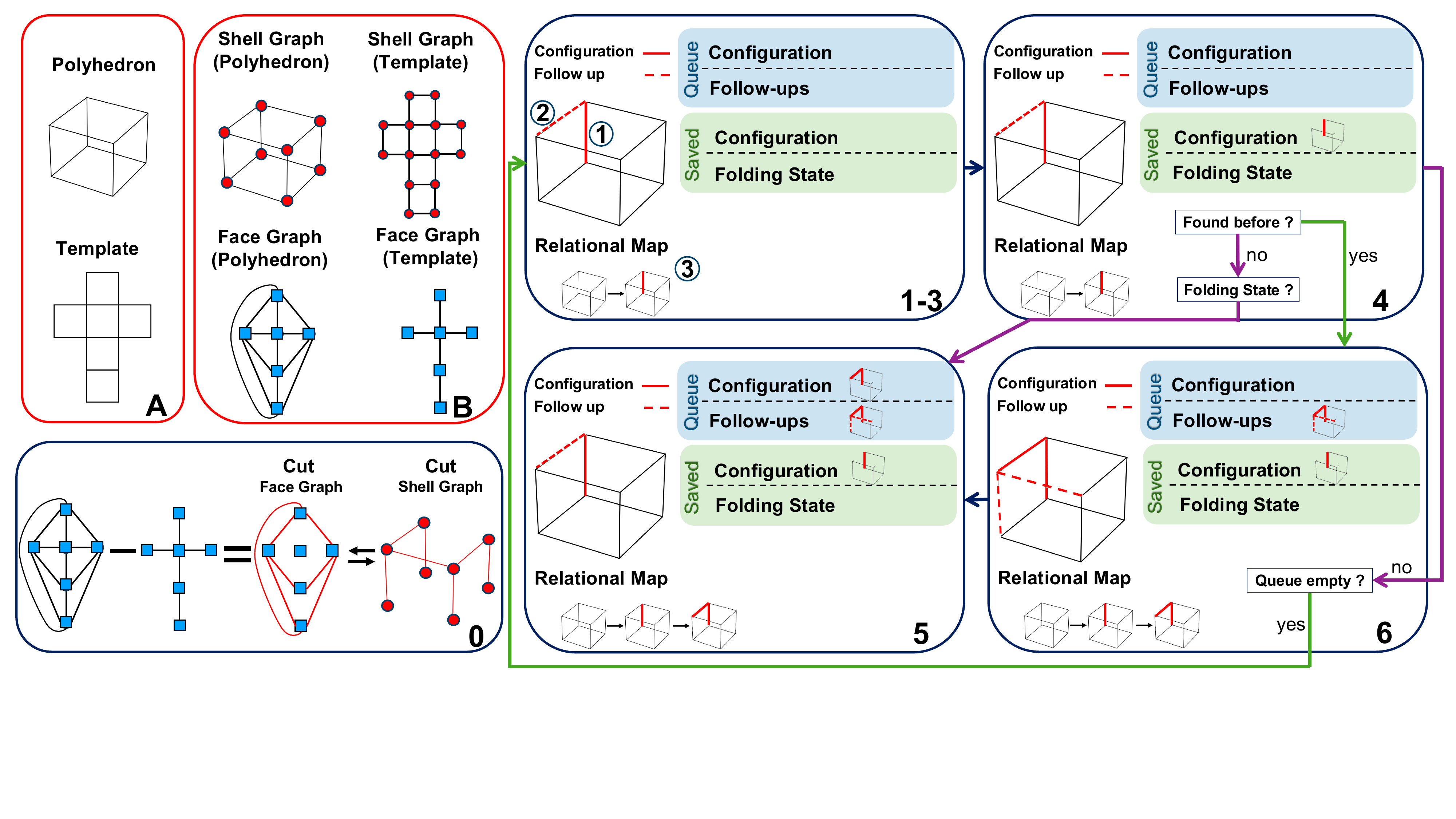}
        \caption{\textbf{Scheme of the algorithm.} The folded polyhedron (cube) and its unfolded template (Latin cross) \textbf{(A)} serve as the starting point of the algorithm. Based on these structures, we construct the shell graph and the face graphs of both the template and the folded polyhedron \textbf{(B)}. These three structures act as inputs to the algorithm. The process begins by constructing the face graph of the cut, obtained by subtracting the face graph of the template from the face graph of the folded polyhedron. This step also yields the shell graph of the cut \textbf{(0)}. The algorithm selects an initial link (edge) on the shell graph, defining the first configuration \textbf{(1)}. Adjacent links to this starting link are identified as potential follow-ups \textbf{(2)}. The folded polyhedron is assigned as the parent configuration of the starting-link configuration \textbf{(3)}. The algorithm checks whether this configuration has been encountered before. If it has, the algorithm moves on to \textbf{6}; else, it is saved as a new configuration \textbf{(4)}. If the configuration is new, the next configurations to evaluate are generated by adding the identified follow up links to the current configuration. These configurations, along with their respective follow-ups, are added to a queue. Each new configuration is designated as a child of the current one in the relational map \textbf{(5)}. A new configuration is retrieved from the queue, and the process returns to step \textbf{(4)}. When the queue is exhausted, the algorithm loops back to \textbf{(1)} and selects a new starting link. If no new links remain and the queue is empty, the algorithm terminates.
        }
        \label{algorithm_explanation}
\end{figure}

To describe the proposed algorithm, consider the case of a cube and its corresponding Latin cross template (Fig.~\ref{algorithm_explanation}.A). The associated \textit{shell} and \textit{face} graphs are shown in Fig.~\ref{algorithm_explanation}.B. Unfolding the cube by cutting along its edges corresponds to removing links in the face graph. The set of all removed links (or edges) forms the cut graph, a subgraph of the face or shell graph that maintains the same nodes but with fewer links. Specifically, the cut graph of the face graph represents the difference between the face graphs of the folded shape and of its unfolded template, while the cut graph of the shell graph corresponds to the set of edges that must be cut along on the cube to fully unfold it (see Fig.~\ref{algorithm_explanation}.0).

\begin{figure}[t]
    \centering
    \includegraphics[width=\textwidth]{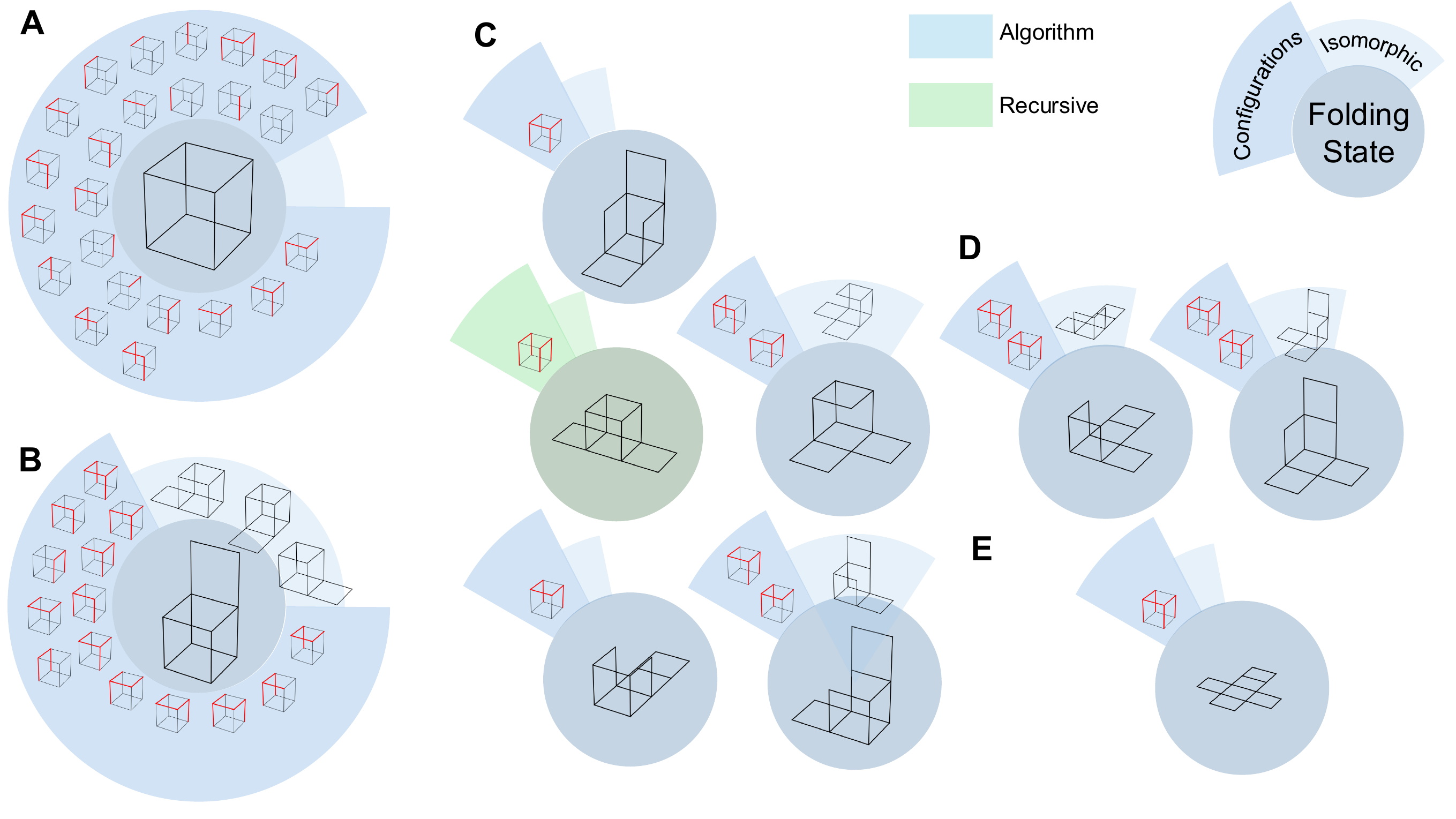}
    \caption{\textbf{Catalog of folding states and configurations for the cube and Latin cross.} Two configurations differ in the set of cut edges, while folding states differ in the set of foldable edges. \textbf{(A)} There are 24 configurations of the cube where no face can be unfolded (i.e., zero foldable edges), despite having certain edges cut (marked in red). \textbf{(B)} When one foldable edge is present, there are 14 distinct configurations, corresponding to four folding states. However, these four folding states are isomorphic, so they are grouped into a single folding state. \textbf{(C)–(E)} Scheme of the folding states, their isomorphisms, and the corresponding configurations for cases with two, three, and five foldable edges. Notably, no folding state exists with exactly four foldable edges. In \textbf{(C)} we distinguish between folding states obtained through a connected path of cut edges (in blue) and those arising from a combination of disconnected cuts (in green), these folding states are obtained by recursively applying the algorithm. \label{confings_foldstates}
    }
\end{figure}

Once we define the initial and final face graphs, to identify all intermediate folding states, we must determine all possible combinations of link removals (edge cuts) that produce valid unfolding steps. However, removing a single link does not always allow a face to unfold. As an example, Fig.~\ref{confings_foldstates}.A shows the 24 configurations obtained by cutting edges of the cube (marked in red) that do not lead to a new folding state, as no face can be unfolded. To systematically identify all intermediate folding states, we begin by choosing a random link from the cut graph and removing it (cutting along the edge) (Fig.~\ref{algorithm_explanation}.1). All adjacent links on the shell cut graph (those sharing a node with the removed link) are then added to a list of candidates for the next removal (Fig.~\ref{algorithm_explanation}.2), and the resulting graph is stored as a child configuration on the relational map (Fig.~\ref{algorithm_explanation}.3). This new configuration is not necessarily a new folding state. Nevertheless, we retain all configurations to reconstruct the full folding sequence and ensure a complete pathway analysis.

After steps (1–3), we check whether the current configuration has been visited before. To do this efficiently, we encode each configuration as a binary number, where each digit corresponds to a specific link in the cut graph: set to zero if the link has been removed and one otherwise. This binary representation uniquely identifies each configuration (see examples in \textit{Supporting Information} \ref{A2_ConfigID}). If the configuration is new, we determine whether it represents a distinct intermediate folding state, meaning that at least one additional face can unfold compared to its parent configuration. If not, it is the same folding state. The edges along which unfolding can occur correspond to specific links in the face graph (known as red links \cite{Coniglio1981, Herrmann_1984}) that, when removed, cause the graph to split into two disconnected components (see \textit{Supporting Information} \ref{A1_RedLinks}). Thus, a new folding state is characterized by an increase in the number of these foldable edges relative to the previous one. Figure~\ref{confings_foldstates} shows all folding states and corresponding configurations for (a) zero, (b) one, (c) two, (d) three, and (e) five foldable edges. As shown in Fig.~\ref{algorithm_explanation}.4, newly identified configurations are added to a list, and the process advances to step 5; otherwise, it proceeds to step 6.

In step 5 (see Fig.~\ref{algorithm_explanation}.5), new configurations are generated by selecting the next adjacent link from the cut graph, based on the list obtained in step 2. The number of possible new configurations corresponds to the number of adjacent links. These new configurations are added to a queue, each paired with its corresponding list of adjacent links associated with the newly selected link (as in step 2). These configurations are labeled as children of the previous one in the relational map, where each connection defines a path that only needs to be explored once, significantly improving the efficiency of the algorithm. In step 6 (Fig.~\ref{algorithm_explanation}.6), we iterate through every configuration in the queue and repeat the process starting from step 4. Once the queue is empty, we return to step 1 (Fig.~\ref{algorithm_explanation}.1) and choose a different starting edge, continuing until all edges are selected.

So far, we have only identified sets of contiguous removed links, leading to pathways that require a connected path of cut edges in the shell graph (blue states in Fig.~\ref{confings_foldstates}.C). However, a shape can also be unfolded by cutting along in different, non-adjacent regions (green states in Fig.~\ref{confings_foldstates}.C). 
These pathways can be obtained by applying the algorithm to the newly found folding states with each folding state being considered as a targeted shape for the algorithm. New folding states found by recursively applying the algorithm will be unfolded in the same way. When all folding states have been identified, the algorithm stops.

The result is a set of configurations, corresponding to subgraphs of the shell graph, grouped into intermediate folding states. The relational map is a directed network of different configurations, with the folded shape on one end and the unfolded template on the other. If we only consider the folding configurations, the network corresponds to the discrete folding landscape, where each node represents a folding state, and the links between nodes indicate possible connections through a sequence of edge cuts. The connection between folding configurations is obtained by going backwards and grouping configurations that do not lead to a new folding state.

At the end, several intermediate states are isomorphic to each other. For example, in the case of the Latin cross, there are four identical folding configurations where all but one face is folded (Fig.~\ref{confings_foldstates}). We group all isomorphic configurations into one, simplifying the discrete landscape. To detect isomorphic states, we analyze the face graphs of different configurations and apply the VF2++ algorithm \cite{Juttner2018}, implemented in the NetworkX package \cite{SciPyProceedings_11}. Note that different templates may share the same face graph; therefore, to distinguish them, we must also check for geometric differences. Once we have the entire landscape for a certain template, we can incorporate all the other known templates to obtain the full discrete folding landscape. For instance, in the case of the cube, in addition to the Latin cross, there are ten other possible templates. The folding landscape for the cube is shown in Fig.~\ref{cube_map}, with the paths corresponding to the Latin cross highlighted in red. Notably, several folding states are common across different templates. Thus, when searching for intermediate configurations and folding states, it is computationally efficient to also consider those obtained from other templates. In the \textit{Supporting Information \ref{A4_TimeEfficiency}}, we demonstrate that for very large shapes, the number of newly discovered configurations and states increases sublinearly with the number of added templates.

\section{Results and Discussion}\label{sec2}

\begin{figure}[!htb]
        \centering
        \includegraphics[width=\textwidth]{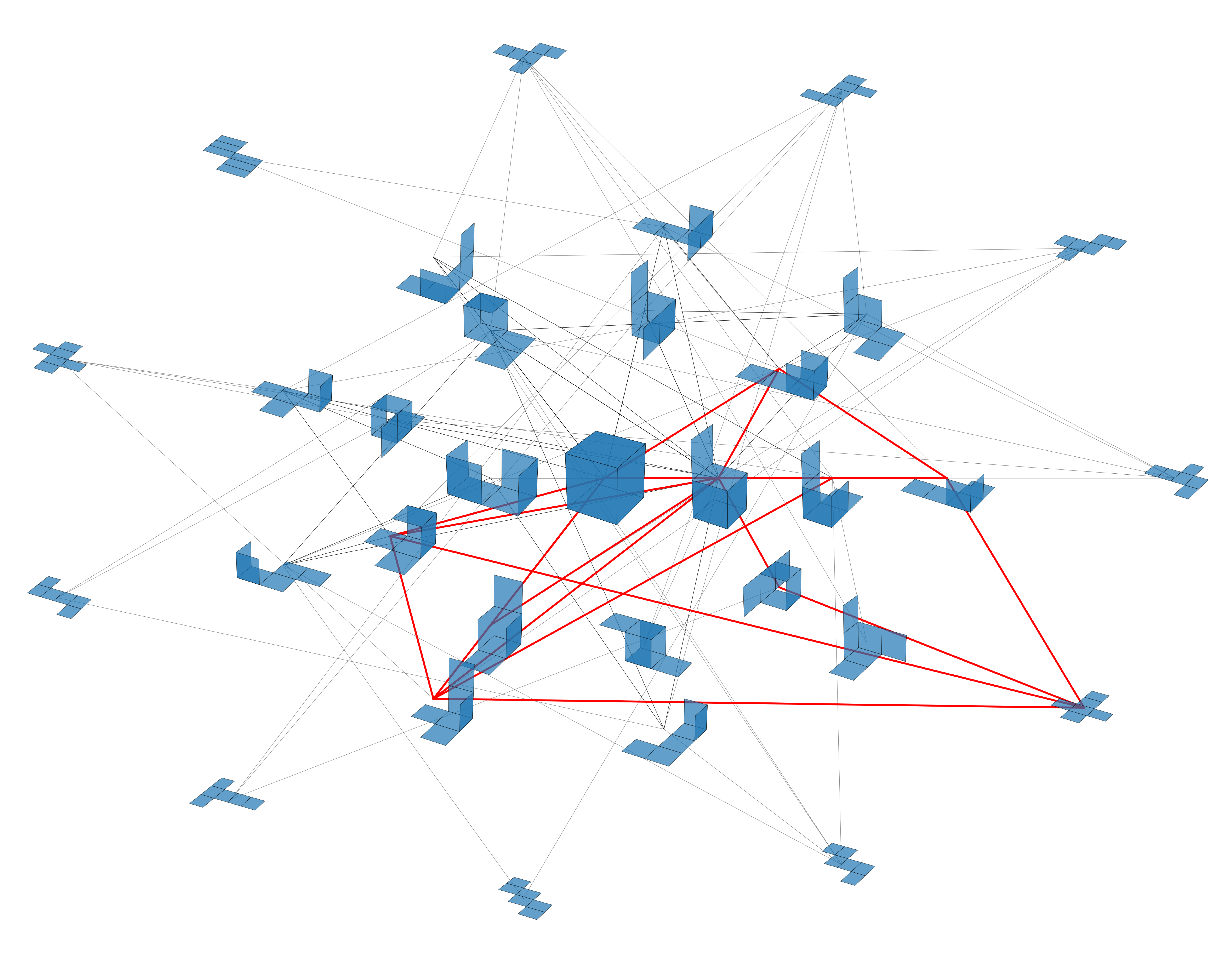}
        \caption{\textbf{Folding landscape for the cube.} The algorithm retrieves a network where each node represents a non-isomorphic folding state, and the links indicate possible transitions between states through a sequence of edge cuts. The cube has eleven distinct templates. The structure in the center (cube) has no red links and the templates have five, structures with the same number of red links are organized in concentric circles with ascending order by their number of red links. The folding pathways for the Latin cross are highlighted in red. \label{cube_map}}
\end{figure}

Figure~\ref{cube_map} shows the folding landscape for the cube with its eleven templates and eighteen intermediate folding states. This network includes all possible non-isomorphic folding states (nodes) and the transitions between them via a sequence of edge cuts (links). The algorithm also retrieves the configurations corresponding to each state (see Fig.~\ref{confings_foldstates}) and the sequence of edge cuts for each link, providing a detailed description of the folding process.

The analysis of this landscape reveals all possible pathways and connections between the folding states. Each template offers multiple pathways to reach the fully folded cube, with the relative statistical weight depending on the folding mechanism. At the microscale, folding can be driven by thermal motion, where the sequence of folds is determined by the rate at which two faces bind to each other~\cite{pandey2011algorithmic, Melo2020}. At the macroscale, folding is typically controlled by actuators or stress relaxation~\cite{Sussman2015, Felton2014}, allowing pathways to be designed either to minimize the number of folds or to reduce the use of actuators.

\begin{figure}
    \centering
    \includegraphics[width=\textwidth]{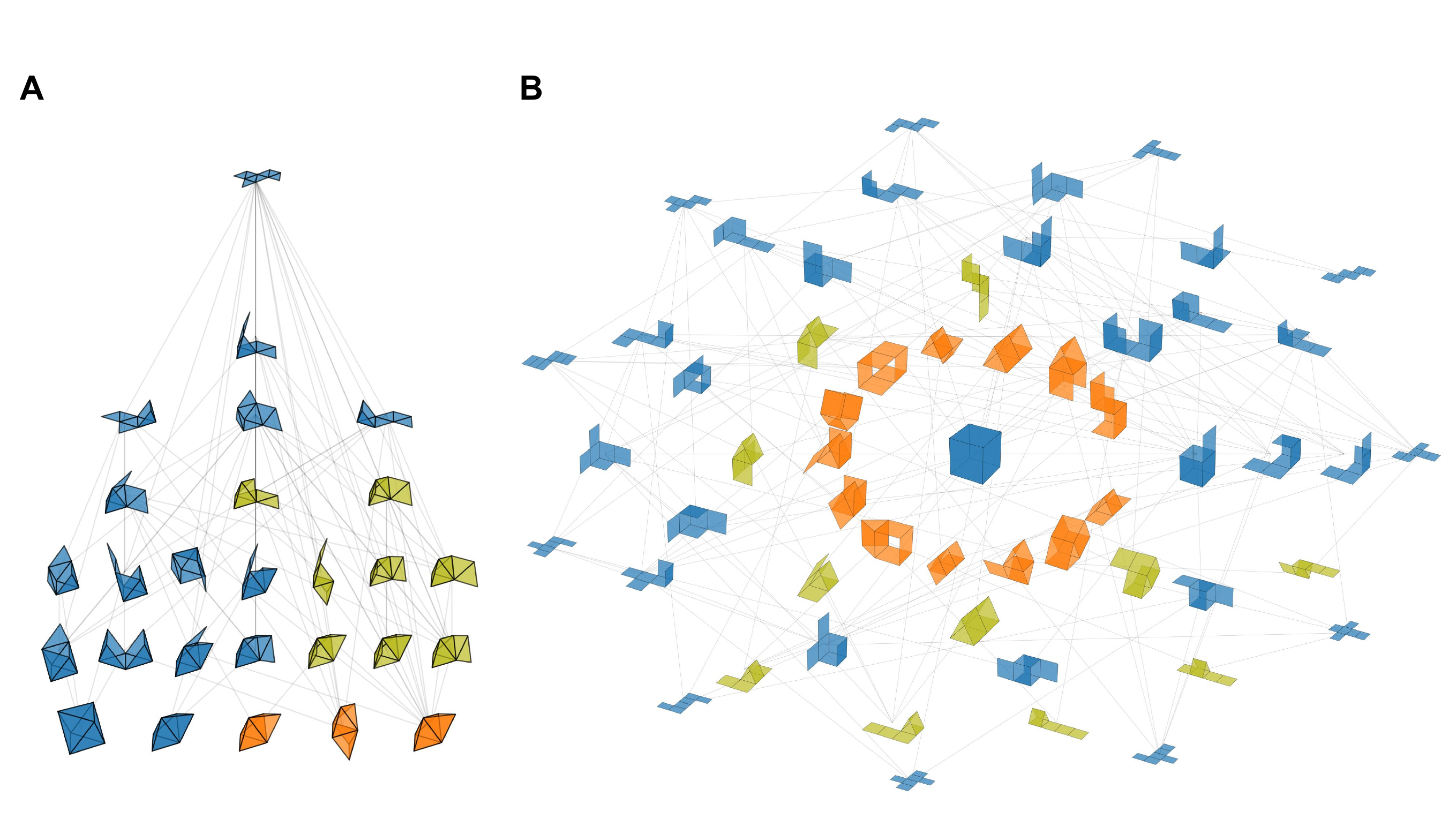}
    \caption{\textbf{Folding landscape for the cube and a multifarious template with misfolds.} Similarly to Figures \ref{cube_map} the network of templates, target structures and intermediate folding states is here shown with additional misfolded structures (in yellow and orange) as well. On \textbf{A} we analyze a multifarious template capable of folding into either an octahedron or a tritetrahedral (boat) structure. A multifarious template is one that can fold into more than one closed structure. To construct the folding landscape, we apply the proposed algorithm, starting from both closed configurations of the same template. Structures are organized by their number of red links starting from the top (the template with seven) to the bottom (the target structures with zero). To the octahedron and the tritetrahedral (boat) structure we add three additional misfolds to which the template can evolve to. These structures in orange with no red links were added as the target structure input to the algorithm together with their respective template. Intermediate folding states which only evolve towards misfolds are shown in yellow. On \textbf{B} several templates and the target structure, the cube, are joined by misfolded folding states added as target structures to the algorithm (orange structures).}
    \label{bothmisfold}
\end{figure}

Knowing the folding landscape is also essential for the design of multifarious templates, which can fold into multiple structures~\cite{Murugan15}. These templates hold promise for the development of shape-changing materials capable of adapting their form in response to external stimuli. Obtaining their folding landscape is straightforward; we apply the algorithm to all possible closed structures. Figure~\ref{bothmisfold}.A (blue structures) presents an example of a template that can fold into either an octahedron or a tritetrahedral (boat) configuration. Identifying possible pathways and critical folds that lead to different final structures can inform the design of the folding process and provide insight into how the final configuration depends on temperature in thermally driven folding~\cite{Pinto2024}.


Up to this point, we have assumed that the final folded state is always a closed polyhedron. However, some folded states may correspond to misfolds, which are partially folded structures that cannot fully assemble into the intended polyhedron without additional edge cuts. These misfolds can be identified through tedious combinatorial analysis or detected dynamically in direct simulations or experiments. Once identified, they must be incorporated into the algorithm as inputs to construct a more extensive folding landscape (Fig.~\ref{bothmisfold}). By doing so, we obtain a comprehensive representation of the folding landscape, including misfolded states, as shown in \textit{Supplementary Material}~\ref{A3_Misfolds} for the two cases considered here: the cube and the multifarious template.

\section{Conclusion}\label{sec13}

We introduced a graph-based algorithm for mapping the complete discrete folding landscape of three-dimensional (3D) structures from two-dimensional (2D) templates consisting of panels connected by flexible hinges. By representing polyhedral templates as graphs of connected faces and edges, our approach efficiently identifies all possible intermediate folding configurations, folding states, and transition pathways. This method overcomes the factorial growth limitations of traditional combinatorial approaches, providing a computationally feasible solution for analyzing complex folding processes.

A key advantage of this framework is its ability to capture not only valid folding sequences but also misfolds, which are states that cannot fully assemble into the intended structure without additional edge cuts. By incorporating these misfolds into the analysis, the algorithm provides a comprehensive representation of the folding landscape, including alternative pathways to correct folding errors. The approach also extends to multifarious templates, which can be folded into multiple distinct closed structures.

The insights gained from this folding landscape have direct implications for self-folding materials, origami-based robotics, and deployable structures. At the microscale, where folding is driven by thermal motion, understanding the sequence of binding interactions can help optimize assembly kinetics~\cite{McMullen2022, pandey2011algorithmic, Dodd2018, Kim2023}. At the macroscale, where actuators and mechanical constraints govern folding, the landscape provides a means to design efficient folding pathways that minimize energy expenditure or mechanical complexity~\cite{Felton2014}. The type and length scale of the system will have direct consequences on its dynamics, these specific details are captured in the distribution of weights for links in the obtained network of folding. However, the network itself is not dependent on the scale and can be studied solely from an abstract point of view.

Beyond its immediate applications, our method lays the foundation for machine learning-driven folding design. The dataset generated from our algorithm comprises folding pathways, misfolded states, and transition probabilities, that can be used to train models that predict novel and optimal folding strategies. This has potential applications in the development of programmable meta-materials, reconfigurable devices, and adaptive structures.

Finally, we demonstrate that as more templates are added, the number of newly discovered configurations and folding states grows sublinearly, suggesting that our approach remains computationally efficient even for large and complex structures. Future work will explore how external forces, geometric constraints, and stochastic effects influence folding landscapes, further refining our understanding of how foldable structures can be designed, controlled, and optimized across scales and disciplines.

\backmatter

\bmhead{Supplementary information}

The code for the algorithm is available on \url{https://github.com/jneves-cftc/kirigami}.

\bmhead{Acknowledgements}

We acknowledge financial support from the Portuguese Foundation for Science and Technology (FCT) under the contracts: UIDB/00618/2020 (DOI:10.54499/UIDB/00618/2020), UIDP/00618/2020 (DOI:10.54499/UIDP/00618/2020), and 2023.00707.BD.

\pagebreak

\begin{appendices}

\section{Red links}\label{A1_RedLinks}
\renewcommand{\thefigure}{A\arabic{figure}}     
\renewcommand{\theHfigure}{A\arabic{figure}}    
\setcounter{figure}{0}

\begin{figure}[!htb]
    \centering
    \includegraphics[width=\textwidth]{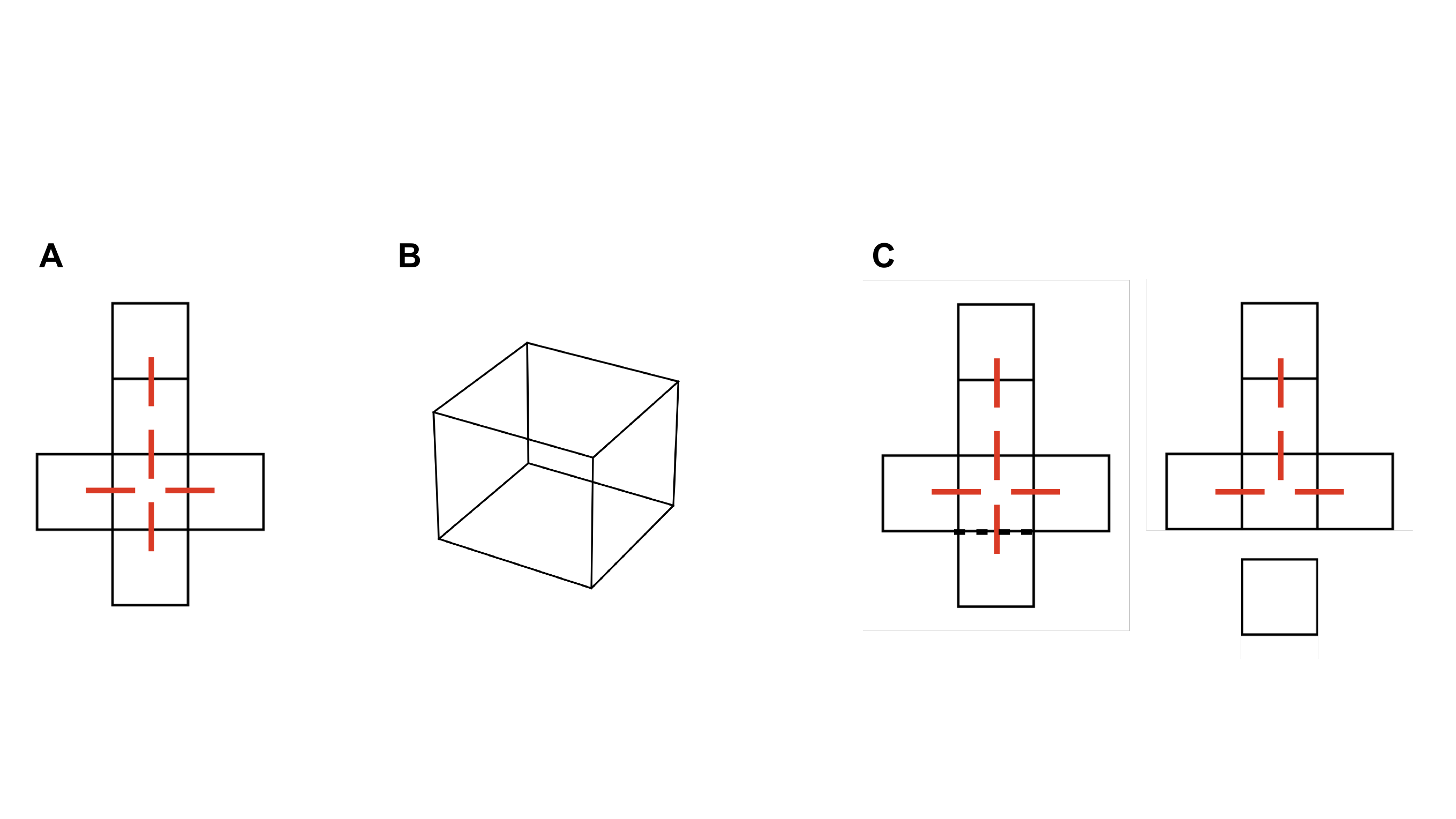}
    \caption{\textbf{Red links on the template and folded structure.} Red links are used to determine which configurations correspond to folding states. The latin cross template of the cube \textbf{A} has 5 red links, these correspond to links on the face graph which when cut separate the graph into two unconnected components as shown on \textbf{C}. The cube \textbf{B} on the other hand has 0 red links, there is no unique edge which when cut separates this structure into two.}
    \label{fig_red_links}
\end{figure}

In graph theory, a red link is a link that, when removed separates the structure into two distinct components. The structure on Fig.~\ref{fig_red_links}.A has five red links whereas structure \ref{fig_red_links}.B has none. If one link from the face graph of structure \ref{fig_red_links}.A is removed, i.e. to cut along any edge in the template, the structure splits into two distinct components, as shown in Fig.~\ref{fig_red_links}.C. On the other hand, cuts along any edge of the cube do not split the structure.

\begin{figure}[!htb]
    \centering
    \includegraphics[width=\textwidth]{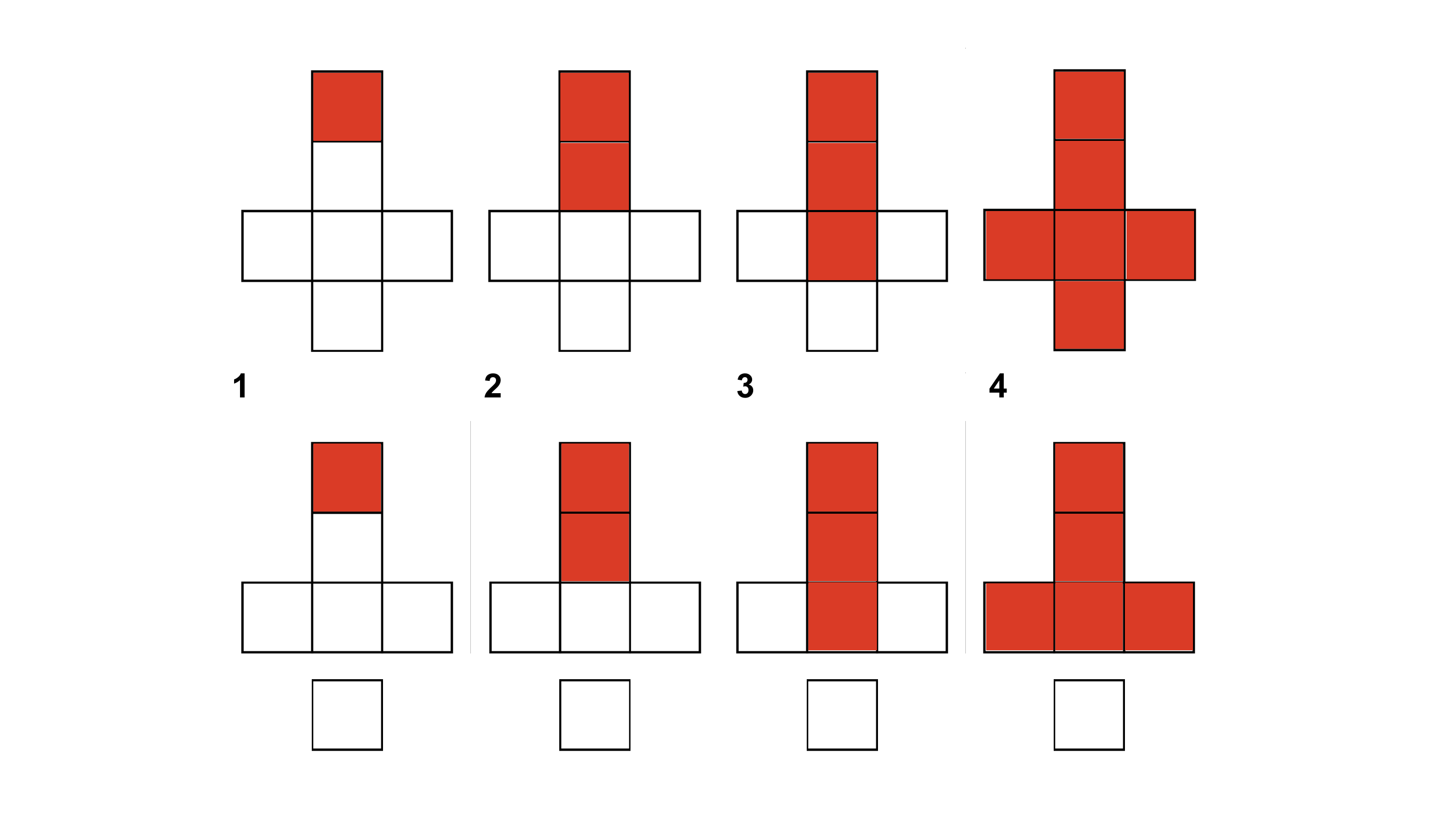}
    \caption{\textbf{Burning algorithm.} Red links are found using a burning algorithm. The algorithm starts by choosing a node to \textit{burn} \textbf{1}, on the following iterations the neighbors of the burnt node will also be burnt \textbf{2}, \textbf{3}, \textbf{4}. The algorithm ends when there are no neighbors left to burn. The structure on top is one piece, thus its nodes (or faces) are all burned in the end. The structure on the bottom corresponds to two separate components, so not all nodes will be burned in the end.}
    \label{fig_burning_alg}
\end{figure}

The number of red links in a specific structure is obtained by applying a burning algorithm \cite{Herrmann_1984}. In Fig.~\ref{fig_burning_alg} we show the employment of this algorithm on the two structures of Fig.~\ref{fig_red_links}.C. The algorithm starts by choosing a node to \textit{burn}, as we are dealing with the face graph of the structure, nodes correspond to faces. On the following iterations, the fire spreads to the neighbors of the burnt nodes, that is, their adjacent faces. The process ends when there are no neighbors left to burn. In the end if all nodes have been burned, the structure is still one piece, if not, the structure is made of two separate components.

To know if a certain link is a red link, we remove it from the face graph and apply this algorithm. If after the link is removed the structure is still one piece, the link is not a red link. If the structure is made of two separate components, the removed link is a red link. 

\section{Identifying Configurations}\label{A2_ConfigID}
\renewcommand{\thefigure}{B\arabic{figure}}
\renewcommand{\theHfigure}{B\arabic{figure}}
\setcounter{figure}{0}

\begin{figure}[!htb]
    \centering
    \includegraphics[width=\textwidth]{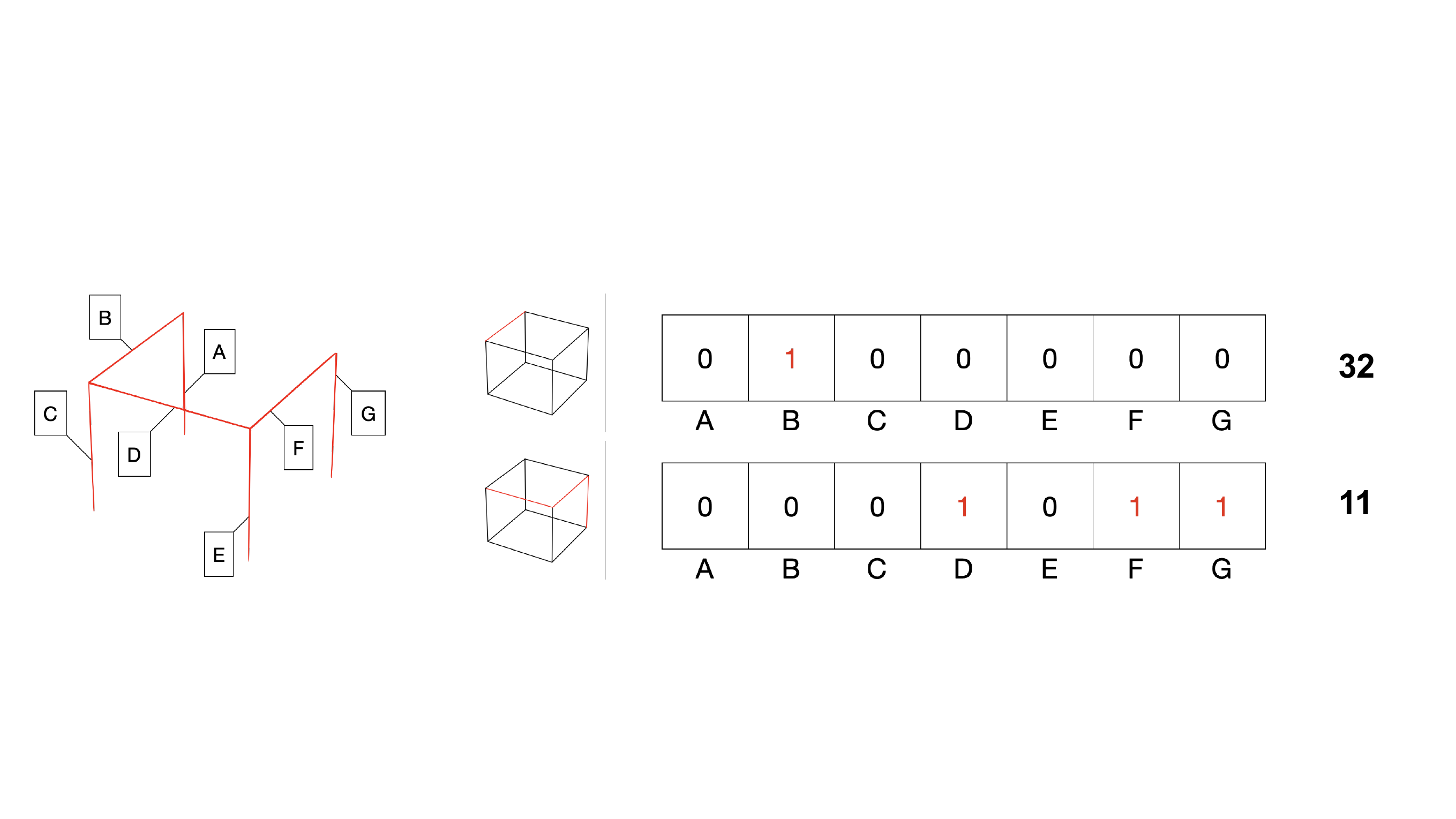}
    \caption{\textbf{Identification of configurations.} Configurations are uniquely identified by a binary number. The edges on the shell cut graph (left) are labeled and ordered, in this case with letters from A to G. Binary numbers for each configuration are constructed by pairing 1's and 0's with the letters that identify the edge, 1 if that edge is on the configuration and 0 otherwise. The top configuration, is a one edge configuration where only edge B is present, thus the binary number $0100000$ (32 in decimal) uniquely identifies this structure. The bottom configuration is composed of edges D, F and G, following the same process the number $0001011$ is constructed (11 in decimal).
    \label{fig_config_id}}
\end{figure}

For the sake of efficiency, all configurations are uniquely identified by a binary number. This increases the speed when inspecting whether a configuration has been found by the algorithm before. The system of identification works by labeling and ordering each edge on the cut graph. In that same order a number is given to each edge, 1 if this edge appears on the configuration and 0 otherwise (Fig.~\ref{fig_config_id}). The sequence of 1's and 0's will thus uniquely identify the configuration. 

\section{Misfolds}\label{A3_Misfolds}
\renewcommand{\thefigure}{C\arabic{figure}}
\renewcommand{\theHfigure}{C\arabic{figure}}
\setcounter{figure}{0}

The input to the algorithm does not need to be a template and a folded polyhedron. Any structure obtained through the process of folding can be used as input to the algorithm, for example a misfold. Misfolds are defined as structures in which a bond needs to be removed in order to correctly fold it. 

On Fig.~\ref{bothmisfold}.B we show the folding landscape of the cube for every possible template as we did on Fig.~\ref{cube_map} (in blue), this time we added misfolded states as target structures (in orange) and their intermediate folding states (in yellow) as well. These misfolded states were obtained by careful observation of a macro-meter scale kirigami when subject to different folding and binding events. These misfolded states with zero red links were used as the final configuration inputs to the algorithm as well as their respective initial configurations (templates). After this, we search for isomorphic structures among the whole catalog of found folding states and construct the map of folding exactly as we did before. In this way, we are able to see the whole evolution of the folding process until it reaches a misfolded state, these states (yellow) are states which can no longer evolve through folding alone to the correctly folded structure which is the cube. 


On Fig.~\ref{bothmisfold}.A we present the folding map of the multifarious configurations (in blue), we have also added three misfolded configurations as inputs (in orange). Similarly to the previous figure we searched for isomorphic folding states and grouped them into one, the resulting map gives information on the key cuts and key binds after which the structure can no longer evolve towards the correctly folded structure.


\section{Efficiency}\label{A4_TimeEfficiency}
\renewcommand{\thefigure}{D\arabic{figure}}
\renewcommand{\theHfigure}{D\arabic{figure}}
\setcounter{figure}{0}



\begin{figure}[!htb]
    \centering
    \includegraphics[width=\textwidth]{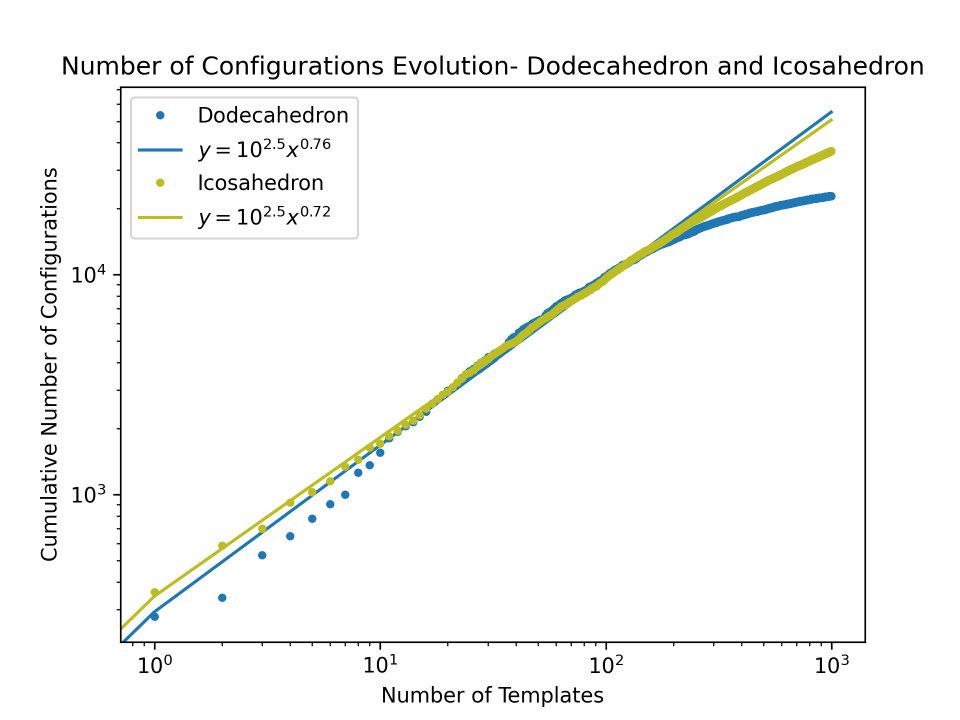}
    \caption{\textbf{Dependency of the cumulative number of configurations with the number of templates.} Scaling of the newly found intermediate configurations with the number of analyzed templates of the two most complex platonic solids, the icosahedron and the dodecahedron, both with more than 48 thousand templates each. The number of newly found configurations scales, at first sight, sublinearly with the number of analyzed templates. In the end a deceleration is observed, indicative of a stagnation in the rate of newly found configurations. The configurational space seems to already be well-probed after a certain magnitude of templates are analyzed.}
    \label{cumulative_conf}
\end{figure}

The algorithm was also applied to the Icosahedron and the Dodecahedron, two of the most complex Platonic solids which have 43380 templates each. To find all possible configurations for the unfolding of one of these polyhedra is a computationally costly process but a possible one. Along with the thousands of configurations found, the rate of discovery for new configurations was also studied regarding the number of templates given to the algorithm. The idea was to construct the folding map of several templates and then combine them one by one, searching and eliminating equivalent configurations as explained before. On Fig.~\ref{cumulative_conf} we can see that the number of newly found configurations scales sub-linearly with the number of added templates. This is an indication that the algorithm is efficient at probing the configurational space and that many common configurations exist in the folding process of different templates. We can also observe that the rate of newly found configurations starts to decrease at a certain point close to a thousand templates. This is an indication that something similar to a plateau is to be reached, and the configurational space has already been mostly explored, with new configurations making fewer and fewer contributions to the configurational space. 

\begin{figure}
    \centering
    \includegraphics[width=\textwidth]{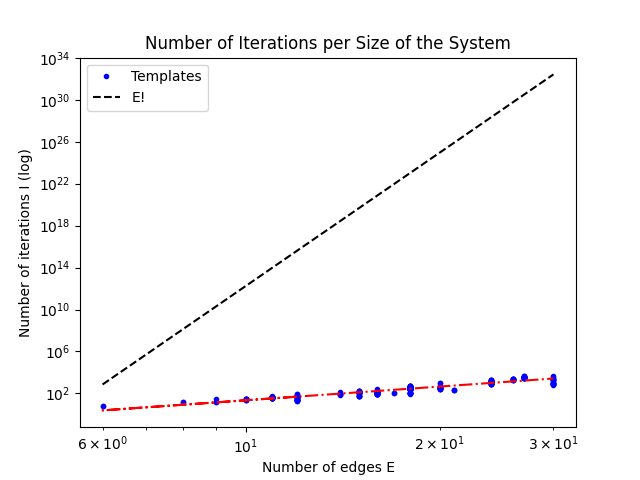}
    \caption{\textbf{Dependency of the number of iterations of the algorithm with the size of the system.} Comparison between the scaling of the number of iteration of the algorithm for different templates of variable size and the factorial explosion of most combinatorial methods. Templates of different polyhedra were used as inputs to the algorithm and the number of iterations of the algorithm was computed against the size of the polyhedra, ie the number of edges. Our algorithm significantly decreases the number of iterations needed to probe the complete folding landscape comparing to simple combinatorial exploration. }
    \label{iterations_edges}
\end{figure}

Another measure of efficiency is the number of iterations the algorithm takes to find the complete folding landscape for a given template and its target structure. If one was to verify all possible intermediate configurations made only by combining different edges to be cut, the pool of configurations to verify would scale factorially with the number of cuts needed to open the structure. As the number of cuts is proportional to the size of the system, we plot on Fig.~\ref{iterations_edges} the factorial scaling and the number of iterations of the algorithm for different templates of polyhedra with variable size. The number of iterations needed to probe the whole folding landscape scales slower than the factorial scaling. This means that the pool of structures to verify is much less with our algorithm than it would be with simple combinatorial methods.

\section{Dodecahedron}\label{A5_Dodecahedron}
\renewcommand{\thefigure}{E\arabic{figure}}
\renewcommand{\theHfigure}{E\arabic{figure}}
\setcounter{figure}{0}

\begin{figure}
    \centering
    \includegraphics[width=0.8\textwidth]{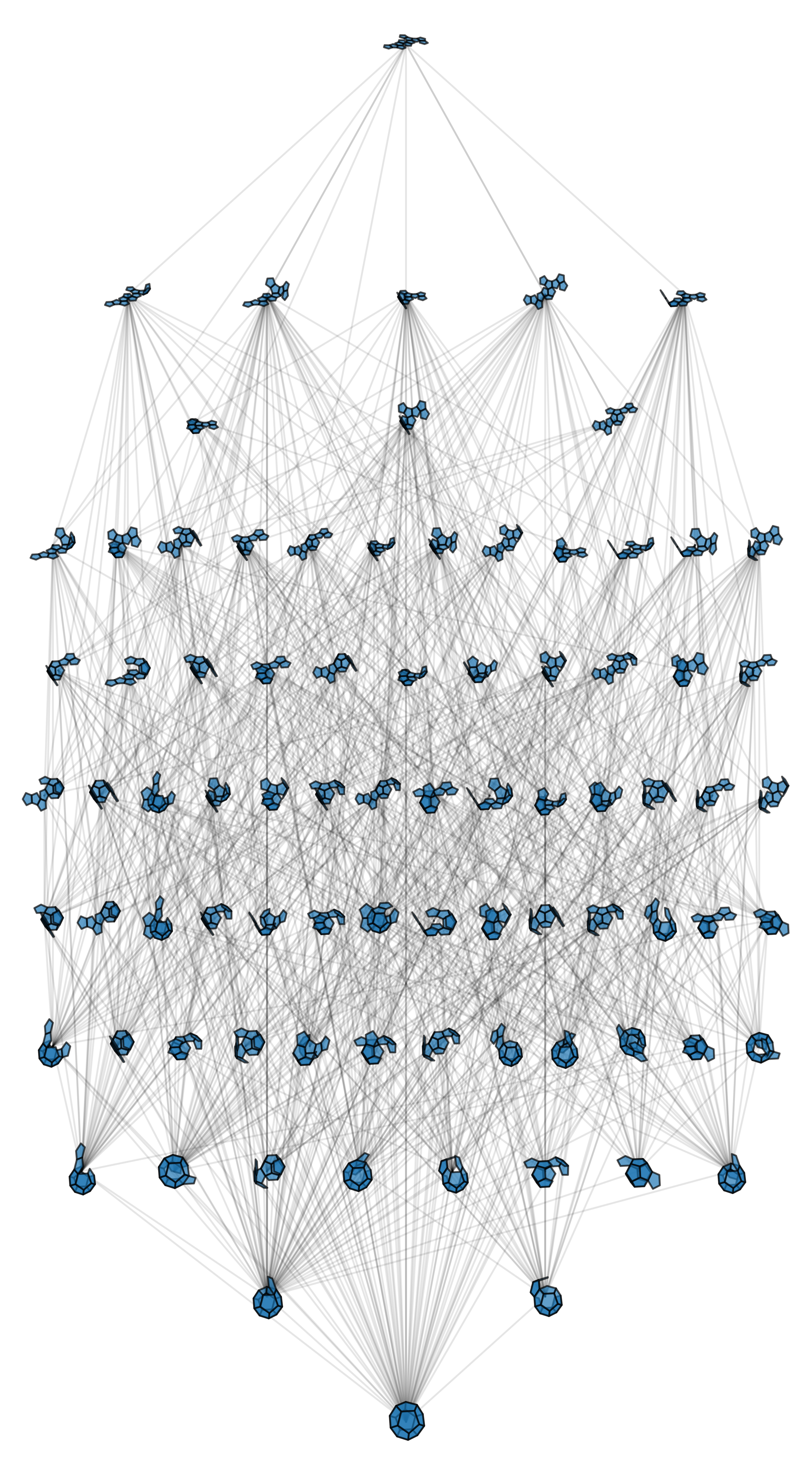}
    \caption{\textbf{Folding landscape for one template of the dodecahedron.} As in Figures \ref{cube_map} and \ref{bothmisfold}, the network of intermediate folding states is here shown for a specific template of the dodecahedron, its target structure. The resulting folding network is much more complex than the previously shown due to the increase on the number of edges of the target structure and consequently on the cut graph. Despite the complexity of the polyhedron we are still able to probe its folding landscape with success.}
    \label{dodecahedron}
\end{figure}

On Fig.~\ref{dodecahedron} the folding network for one template of the dodecahedron is shown. 

\end{appendices}



\begin{thebibliography}{10}
\expandafter\ifx\csname url\endcsname\relax
  \def\url#1{\burl{#1}}\fi
\expandafter\ifx\csname urlprefix\endcsname\relax\def\urlprefix{URL }\fi
\providecommand{\bibinfo}[2]{#2}
\providecommand{\eprint}[2][]{\url{#2}}
\providecommand{\doi}[1]{\url{https://doi.org/#1}}
\bibcommenthead

\bibitem{meta_materials_review1}
\bibinfo{author}{Qi, J.} \emph{et~al.}
\newblock \bibinfo{title}{Recent progress in active mechanical metamaterials
  and construction principles}.
\newblock \emph{\bibinfo{journal}{Advanced Science}}
  \textbf{\bibinfo{volume}{9}}, \bibinfo{pages}{2102662}
  (\bibinfo{year}{2022}).
\newblock
  \urlprefix\url{https://onlinelibrary.wiley.com/doi/abs/10.1002/advs.202102662}.

\bibitem{meta_materials_review2}
\bibinfo{author}{Choi, G. P.~T.}
\newblock \bibinfo{title}{Computational design of art-inspired metamaterials}.
\newblock \emph{\bibinfo{journal}{Nature Computational Science}}
  \textbf{\bibinfo{volume}{4}}, \bibinfo{pages}{549--552}
  (\bibinfo{year}{2024}).
\newblock \urlprefix\url{https://doi.org/10.1038/s43588-024-00671-y}.

\bibitem{Chen2020}
\bibinfo{author}{Chen, S.}, \bibinfo{author}{Chen, J.}, \bibinfo{author}{Zhang,
  X.}, \bibinfo{author}{Li, Z.-Y.} \& \bibinfo{author}{Li, J.}
\newblock \bibinfo{title}{Kirigami/origami: unfolding the new regime of
  advanced 3d microfabrication/nanofabrication with ``folding''}.
\newblock \emph{\bibinfo{journal}{Light: Science {\&} Applications}}
  \textbf{\bibinfo{volume}{9}}, \bibinfo{pages}{75} (\bibinfo{year}{2020}).
\newblock \urlprefix\url{https://doi.org/10.1038/s41377-020-0309-9}.

\bibitem{Felton2014}
\bibinfo{author}{Felton, S.}, \bibinfo{author}{Tolley, M.},
  \bibinfo{author}{Demaine, E.}, \bibinfo{author}{Rus, D.} \&
  \bibinfo{author}{Wood, R.}
\newblock \bibinfo{title}{A method for building self-folding machines}.
\newblock \emph{\bibinfo{journal}{Science}} \textbf{\bibinfo{volume}{345}},
  \bibinfo{pages}{644--646} (\bibinfo{year}{2014}).
\newblock
  \urlprefix\url{https://www.science.org/doi/abs/10.1126/science.1252610}.

\bibitem{Gu2023}
\bibinfo{author}{Gu, H.} \emph{et~al.}
\newblock \bibinfo{title}{Self-folding soft-robotic chains with reconfigurable
  shapes and functionalities}.
\newblock \emph{\bibinfo{journal}{Nature Communications}}
  \textbf{\bibinfo{volume}{14}}, \bibinfo{pages}{1263} (\bibinfo{year}{2023}).
\newblock \urlprefix\url{https://doi.org/10.1038/s41467-023-36819-z}.

\bibitem{Pedivellano2024}
\bibinfo{author}{Pedivellano, A.} \& \bibinfo{author}{Pellegrino, S.}
\newblock \bibinfo{title}{Folding kinematics of kirigami-inspired space
  structures}.
\newblock \emph{\bibinfo{journal}{International Journal of Solids and
  Structures}} \textbf{\bibinfo{volume}{300}}, \bibinfo{pages}{112865}
  (\bibinfo{year}{2024}).
\newblock
  \urlprefix\url{https://www.sciencedirect.com/science/article/pii/S0020768324002245}.

\bibitem{Zirbel2013}
\bibinfo{author}{Zirbel, S.~A.} \emph{et~al.}
\newblock \bibinfo{title}{Accommodating thickness in origami-based deployable
  arrays1}.
\newblock \emph{\bibinfo{journal}{Journal of Mechanical Design}}
  \textbf{\bibinfo{volume}{135}}, \bibinfo{pages}{111005}
  (\bibinfo{year}{2013}).
\newblock \urlprefix\url{https://doi.org/10.1115/1.4025372}.

\bibitem{mitani2024}
\bibinfo{author}{Mitani, J.}
\newblock \bibinfo{title}{Pillow box design} (\bibinfo{year}{2024}).
\newblock \bibinfo{note}{Preprint at \url{https://arxiv.org/abs/2410.17593}},
  \eprint{2410.17593}.

\bibitem{Zhu2024}
\bibinfo{author}{Zhu, Y.} \& \bibinfo{author}{Filipov, E.~T.}
\newblock \bibinfo{title}{Large-scale modular and uniformly thick
  origami-inspired adaptable and load-carrying structures}.
\newblock \emph{\bibinfo{journal}{Nature Communications}}
  \textbf{\bibinfo{volume}{15}}, \bibinfo{pages}{2353} (\bibinfo{year}{2024}).
\newblock \urlprefix\url{https://doi.org/10.1038/s41467-024-46667-0}.

\bibitem{Babaee2021}
\bibinfo{author}{Babaee, S.} \emph{et~al.}
\newblock \bibinfo{title}{Kirigami-inspired stents for sustained local delivery
  of therapeutics}.
\newblock \emph{\bibinfo{journal}{Nature Materials}}
  \textbf{\bibinfo{volume}{20}}, \bibinfo{pages}{1085--1092}
  (\bibinfo{year}{2021}).
\newblock \urlprefix\url{https://doi.org/10.1038/s41563-021-01031-1}.

\bibitem{Huang2019}
\bibinfo{author}{Huang, H.-W.} \emph{et~al.}
\newblock \bibinfo{title}{Adaptive locomotion of artificial microswimmers}.
\newblock \emph{\bibinfo{journal}{Science Advances}}
  \textbf{\bibinfo{volume}{5}}, \bibinfo{pages}{eaau1532}
  (\bibinfo{year}{2019}).
\newblock
  \urlprefix\url{https://www.science.org/doi/abs/10.1126/sciadv.aau1532}.

\bibitem{McMullen2022}
\bibinfo{author}{McMullen, A.}, \bibinfo{author}{Mu{\~{n}}oz~Basagoiti, M.},
  \bibinfo{author}{Zeravcic, Z.} \& \bibinfo{author}{Brujic, J.}
\newblock \bibinfo{title}{Self-assembly of emulsion droplets through
  programmable folding}.
\newblock \emph{\bibinfo{journal}{Nature}} \textbf{\bibinfo{volume}{610}},
  \bibinfo{pages}{502--506} (\bibinfo{year}{2022}).
\newblock \urlprefix\url{https://doi.org/10.1038/s41586-022-05198-8}.

\bibitem{Veneziano2020}
\bibinfo{author}{Veneziano, R.} \emph{et~al.}
\newblock \bibinfo{title}{Role of nanoscale antigen organization on b-cell
  activation probed using dna origami}.
\newblock \emph{\bibinfo{journal}{Nature Nanotechnology}}
  \textbf{\bibinfo{volume}{15}}, \bibinfo{pages}{716--723}
  (\bibinfo{year}{2020}).
\newblock \urlprefix\url{https://doi.org/10.1038/s41565-020-0719-0}.

\bibitem{Kim2023}
\bibinfo{author}{Kim, M.} \emph{et~al.}
\newblock \bibinfo{title}{Harnessing a paper-folding mechanism for
  reconfigurable dna origami}.
\newblock \emph{\bibinfo{journal}{Nature}} \textbf{\bibinfo{volume}{619}},
  \bibinfo{pages}{78--86} (\bibinfo{year}{2023}).
\newblock \urlprefix\url{https://doi.org/10.1038/s41586-023-06181-7}.

\bibitem{Cells2012}
\bibinfo{author}{Kuribayashi-Shigetomi, K.}, \bibinfo{author}{Onoe, H.} \&
  \bibinfo{author}{Takeuchi, S.}
\newblock \bibinfo{title}{Cell origami: Self-folding of three-dimensional
  cell-laden microstructures driven by cell traction force}.
\newblock \emph{\bibinfo{journal}{PLOS ONE}} \textbf{\bibinfo{volume}{7}},
  \bibinfo{pages}{1--8} (\bibinfo{year}{2012}).
\newblock \urlprefix\url{https://doi.org/10.1371/journal.pone.0051085}.

\bibitem{Lamoureux2015}
\bibinfo{author}{Lamoureux, A.}, \bibinfo{author}{Lee, K.},
  \bibinfo{author}{Shlian, M.}, \bibinfo{author}{Forrest, S.~R.} \&
  \bibinfo{author}{Shtein, M.}
\newblock \bibinfo{title}{Dynamic kirigami structures for integrated solar
  tracking}.
\newblock \emph{\bibinfo{journal}{Nature Communications}}
  \textbf{\bibinfo{volume}{6}}, \bibinfo{pages}{8092} (\bibinfo{year}{2015}).
\newblock \urlprefix\url{https://doi.org/10.1038/ncomms9092}.

\bibitem{Yasuda2015}
\bibinfo{author}{Yasuda, H.} \& \bibinfo{author}{Yang, J.}
\newblock \bibinfo{title}{Reentrant origami-based metamaterials with negative
  poisson's ratio and bistability}.
\newblock \emph{\bibinfo{journal}{Phys. Rev. Lett.}}
  \textbf{\bibinfo{volume}{114}}, \bibinfo{pages}{185502}
  (\bibinfo{year}{2015}).
\newblock
  \urlprefix\url{https://link.aps.org/doi/10.1103/PhysRevLett.114.185502}.

\bibitem{Wang2017}
\bibinfo{author}{Wang, Z.} \emph{et~al.}
\newblock \bibinfo{title}{Origami-based reconfigurable metamaterials for
  tunable chirality}.
\newblock \emph{\bibinfo{journal}{Advanced Materials}}
  \textbf{\bibinfo{volume}{29}}, \bibinfo{pages}{1700412}
  (\bibinfo{year}{2017}).
\newblock
  \urlprefix\url{https://onlinelibrary.wiley.com/doi/abs/10.1002/adma.201700412}.

\bibitem{Fang2018}
\bibinfo{author}{Fang, H.}, \bibinfo{author}{Chu, S.-C.~A.},
  \bibinfo{author}{Xia, Y.} \& \bibinfo{author}{Wang, K.-W.}
\newblock \bibinfo{title}{Programmable self-locking origami mechanical
  metamaterials}.
\newblock \emph{\bibinfo{journal}{Advanced Materials}}
  \textbf{\bibinfo{volume}{30}}, \bibinfo{pages}{1706311}
  (\bibinfo{year}{2018}).
\newblock
  \urlprefix\url{https://onlinelibrary.wiley.com/doi/abs/10.1002/adma.201706311}.

\bibitem{Tao2023}
\bibinfo{author}{Tao, J.}, \bibinfo{author}{Khosravi, H.},
  \bibinfo{author}{Deshpande, V.} \& \bibinfo{author}{Li, S.}
\newblock \bibinfo{title}{Engineering by cuts: How kirigami principle enables
  unique mechanical properties and functionalities}.
\newblock \emph{\bibinfo{journal}{Advanced Science}}
  \textbf{\bibinfo{volume}{10}}, \bibinfo{pages}{2204733}
  (\bibinfo{year}{2023}).
\newblock
  \urlprefix\url{https://onlinelibrary.wiley.com/doi/abs/10.1002/advs.202204733}.

\bibitem{Li2021}
\bibinfo{author}{Li, Y.}, \bibinfo{author}{Zhang, Q.}, \bibinfo{author}{Hong,
  Y.} \& \bibinfo{author}{Yin, J.}
\newblock \bibinfo{title}{3d transformable modular kirigami based programmable
  metamaterials}.
\newblock \emph{\bibinfo{journal}{Advanced Functional Materials}}
  \textbf{\bibinfo{volume}{31}}, \bibinfo{pages}{2105641}
  (\bibinfo{year}{2021}).
\newblock
  \urlprefix\url{https://onlinelibrary.wiley.com/doi/abs/10.1002/adfm.202105641}.

\bibitem{Zheng2014}
\bibinfo{author}{Zheng, X.} \emph{et~al.}
\newblock \bibinfo{title}{Ultralight, ultrastiff mechanical metamaterials}.
\newblock \emph{\bibinfo{journal}{Science}} \textbf{\bibinfo{volume}{344}},
  \bibinfo{pages}{1373--1377} (\bibinfo{year}{2014}).
\newblock
  \urlprefix\url{https://www.science.org/doi/abs/10.1126/science.1252291}.

\bibitem{Guo2021}
\bibinfo{author}{Guo, X.} \emph{et~al.}
\newblock \bibinfo{title}{Designing mechanical metamaterials with
  kirigami-inspired, hierarchical constructions for giant positive and negative
  thermal expansion}.
\newblock \emph{\bibinfo{journal}{Advanced Materials}}
  \textbf{\bibinfo{volume}{33}}, \bibinfo{pages}{2004919}
  (\bibinfo{year}{2021}).
\newblock
  \urlprefix\url{https://advanced.onlinelibrary.wiley.com/doi/abs/10.1002/adma.202004919}.

\bibitem{Li2020}
\bibinfo{author}{Li, Z.}, \bibinfo{author}{Chen, W.}, \bibinfo{author}{Hao,
  H.}, \bibinfo{author}{Yang, Q.} \& \bibinfo{author}{Fang, R.}
\newblock \bibinfo{title}{Energy absorption of kirigami modified corrugated
  structure}.
\newblock \emph{\bibinfo{journal}{Thin-Walled Structures}}
  \textbf{\bibinfo{volume}{154}}, \bibinfo{pages}{106829}
  (\bibinfo{year}{2020}).
\newblock
  \urlprefix\url{https://www.sciencedirect.com/science/article/pii/S0263823120307072}.

\bibitem{Zhang2022}
\bibinfo{author}{Zhang, Q.} \emph{et~al.}
\newblock \bibinfo{title}{Impact behavior of corrugated-core infilling foam
  sandwich composite structure}.
\newblock \emph{\bibinfo{journal}{Case Studies in Construction Materials}}
  \textbf{\bibinfo{volume}{17}}, \bibinfo{pages}{e01418}
  (\bibinfo{year}{2022}).
\newblock
  \urlprefix\url{https://www.sciencedirect.com/science/article/pii/S2214509522005502}.

\bibitem{Fathers2015}
\bibinfo{author}{Fathers, R.}, \bibinfo{author}{Gattas, J.} \&
  \bibinfo{author}{You, Z.}
\newblock \bibinfo{title}{Quasi-static crushing of eggbox, cube, and modified
  cube foldcore sandwich structures}.
\newblock \emph{\bibinfo{journal}{International Journal of Mechanical
  Sciences}} \textbf{\bibinfo{volume}{101-102}}, \bibinfo{pages}{421--428}
  (\bibinfo{year}{2015}).
\newblock
  \urlprefix\url{https://www.sciencedirect.com/science/article/pii/S0020740315003021}.

\bibitem{Sussman2015}
\bibinfo{author}{Sussman, D.~M.} \emph{et~al.}
\newblock \bibinfo{title}{Algorithmic lattice kirigami: A route to pluripotent
  materials}.
\newblock \emph{\bibinfo{journal}{Proceedings of the National Academy of
  Sciences}} \textbf{\bibinfo{volume}{112}}, \bibinfo{pages}{7449--7453}
  (\bibinfo{year}{2015}).
\newblock \urlprefix\url{https://www.pnas.org/doi/abs/10.1073/pnas.1506048112}.

\bibitem{An2020}
\bibinfo{author}{An, N.}, \bibinfo{author}{Domel, A.~G.},
  \bibinfo{author}{Zhou, J.}, \bibinfo{author}{Rafsanjani, A.} \&
  \bibinfo{author}{Bertoldi, K.}
\newblock \bibinfo{title}{Programmable hierarchical kirigami}.
\newblock \emph{\bibinfo{journal}{Advanced Functional Materials}}
  \textbf{\bibinfo{volume}{30}}, \bibinfo{pages}{1906711}
  (\bibinfo{year}{2020}).
\newblock
  \urlprefix\url{https://onlinelibrary.wiley.com/doi/abs/10.1002/adfm.201906711}.

\bibitem{Neville2016}
\bibinfo{author}{Neville, R.~M.}, \bibinfo{author}{Scarpa, F.} \&
  \bibinfo{author}{Pirrera, A.}
\newblock \bibinfo{title}{Shape morphing kirigami mechanical metamaterials}.
\newblock \emph{\bibinfo{journal}{Scientific Reports}}
  \textbf{\bibinfo{volume}{6}}, \bibinfo{pages}{31067} (\bibinfo{year}{2016}).
\newblock \urlprefix\url{https://doi.org/10.1038/srep31067}.

\bibitem{O’Rourke_2011}
\bibinfo{author}{O’Rourke, J.}
\newblock \emph{\bibinfo{title}{How to Fold It: The Mathematics of Linkages,
  Origami, and Polyhedra}}  (\bibinfo{publisher}{Cambridge University Press},
  \bibinfo{year}{2011}).

\bibitem{Demaine_O’Rourke_2007}
\bibinfo{author}{Demaine, E.~D.} \& \bibinfo{author}{O’Rourke, J.}
\newblock \emph{\bibinfo{title}{Geometric Folding Algorithms: Linkages,
  Origami, Polyhedra}}  (\bibinfo{publisher}{Cambridge University Press},
  \bibinfo{year}{2007}).

\bibitem{Dodd2018}
\bibinfo{author}{Dodd, P.~M.}, \bibinfo{author}{Damasceno, P.~F.} \&
  \bibinfo{author}{Glotzer, S.~C.}
\newblock \bibinfo{title}{Universal folding pathways of polyhedron nets}.
\newblock \emph{\bibinfo{journal}{Proceedings of the National Academy of
  Sciences}} \textbf{\bibinfo{volume}{115}}, \bibinfo{pages}{E6690--E6696}
  (\bibinfo{year}{2018}).
\newblock \urlprefix\url{https://www.pnas.org/doi/abs/10.1073/pnas.1722681115}.

\bibitem{Kaplan2014}
\bibinfo{author}{Kaplan, R.}, \bibinfo{author}{Klobušický, J.}, 
  \bibinfo{author}{Pandey, S.}, \bibinfo{author}{Gracias, D.~H.} \&
  \bibinfo{author}{Menon, G.}
\newblock \bibinfo{title}{Building Polyhedra by Self-Assembly: Theory and Experiment}.
\newblock \emph{\bibinfo{journal}{Artificial Life}}
  \textbf{\bibinfo{volume}{20}}, \bibinfo{pages}{409--439}
  (\bibinfo{year}{2014}).
\newblock \urlprefix\url{https://doi.org/10.1162/ARTL_a_00144}.

\bibitem{JohnsonChyzhykov2016}
\bibinfo{author}{Johnson-Chyzhykov, D.} \& \bibinfo{author}{Menon, G.}
\newblock \bibinfo{title}{The Building Game: From Enumerative Combinatorics to Conformational Diffusion}.
\newblock \emph{\bibinfo{journal}{Journal of Nonlinear Science}}
  \textbf{\bibinfo{volume}{26}}, \bibinfo{pages}{815--845}
  (\bibinfo{year}{2016}).
\newblock \urlprefix\url{https://doi.org/10.1007/s00332-016-9291-z}.

\bibitem{Araujo2018}
\bibinfo{author}{Ara\'ujo, N. A.~M.}, \bibinfo{author}{da~Costa, R.~A.},
  \bibinfo{author}{Dorogovtsev, S.~N.} \& \bibinfo{author}{Mendes, J. F.~F.}
\newblock \bibinfo{title}{Finding the optimal nets for self-folding kirigami}.
\newblock \emph{\bibinfo{journal}{Phys. Rev. Lett.}}
  \textbf{\bibinfo{volume}{120}}, \bibinfo{pages}{188001}
  (\bibinfo{year}{2018}).
\newblock
  \urlprefix\url{https://link.aps.org/doi/10.1103/PhysRevLett.120.188001}.

\bibitem{Coniglio1981}
\bibinfo{author}{Coniglio, A.}
\newblock \bibinfo{title}{Thermal phase transition of the dilute $s$-state
  potts and $n$-vector models at the percolation threshold}.
\newblock \emph{\bibinfo{journal}{Phys. Rev. Lett.}}
  \textbf{\bibinfo{volume}{46}}, \bibinfo{pages}{250--253}
  (\bibinfo{year}{1981}).
\newblock \urlprefix\url{https://link.aps.org/doi/10.1103/PhysRevLett.46.250}.

\bibitem{Herrmann_1984}
\bibinfo{author}{Herrmann, H.~J.}, \bibinfo{author}{Hong, D.~C.} \&
  \bibinfo{author}{Stanley, H.~E.}
\newblock \bibinfo{title}{Backbone and elastic backbone of percolation clusters
  obtained by the new method of 'burning'}.
\newblock \emph{\bibinfo{journal}{Journal of Physics A: Mathematical and
  General}} \textbf{\bibinfo{volume}{17}}, \bibinfo{pages}{L261}
  (\bibinfo{year}{1984}).
\newblock \urlprefix\url{https://dx.doi.org/10.1088/0305-4470/17/5/008}.

\bibitem{Juttner2018}
\bibinfo{author}{Jüttner, A.} \& \bibinfo{author}{Madarasi, P.}
\newblock \bibinfo{title}{Vf2++—an improved subgraph isomorphism algorithm}.
\newblock \emph{\bibinfo{journal}{Discrete Applied Mathematics}}
  \textbf{\bibinfo{volume}{242}}, \bibinfo{pages}{69--81}
  (\bibinfo{year}{2018}).
\newblock
  \urlprefix\url{https://www.sciencedirect.com/science/article/pii/S0166218X18300829}.

\bibitem{SciPyProceedings_11}
\bibinfo{author}{Hagberg, A.~A.}, \bibinfo{author}{Schult, D.~A.} \&
  \bibinfo{author}{Swart, P.~J.}
\newblock \bibinfo{editor}{Varoquaux, G.}, \bibinfo{editor}{Vaught, T.} \&
  \bibinfo{editor}{Millman, J.} (eds) \emph{\bibinfo{title}{Exploring network
  structure, dynamics, and function using networkx}}.
\newblock (eds \bibinfo{editor}{Varoquaux, G.}, \bibinfo{editor}{Vaught, T.} \&
  \bibinfo{editor}{Millman, J.}) \emph{\bibinfo{booktitle}{Proceedings of the
  7th Python in Science Conference}}, \bibinfo{pages}{11 -- 15}
  (\bibinfo{address}{Pasadena, CA USA}, \bibinfo{year}{2008}).

\bibitem{pandey2011algorithmic}
\bibinfo{author}{Pandey, S.} \emph{et~al.}
\newblock \bibinfo{title}{Algorithmic design of self-folding polyhedra}.
\newblock \emph{\bibinfo{journal}{Proc. Natl. Acad. Sci. U.S.A.}}
  \textbf{\bibinfo{volume}{108}}, \bibinfo{pages}{19885--19890}
  (\bibinfo{year}{2011}).
\newblock \urlprefix\url{https://doi.org/10.1073/pnas.1110857108}.

\bibitem{Melo2020}
\bibinfo{author}{Melo, H. P.~M.}, \bibinfo{author}{Dias, C.~S.} \&
  \bibinfo{author}{Ara\'ujo, N. A.~M.}
\newblock \bibinfo{title}{{Optimal Number of Faces for Fast Self-Folding
  Kirigami}}.
\newblock \emph{\bibinfo{journal}{Commun. Phys.}} \textbf{\bibinfo{volume}{3}},
  \bibinfo{pages}{154} (\bibinfo{year}{2020}).
\newblock \urlprefix\url{https://doi.org/10.1038/s42005-020-00423-0}.

\bibitem{Murugan15}
\bibinfo{author}{Murugan, A.}, \bibinfo{author}{Zeravcic, Z.},
  \bibinfo{author}{Brenner, M.~P.} \& \bibinfo{author}{Leibler, S.}
\newblock \bibinfo{title}{Multifarious assembly mixtures: Systems allowing
  retrieval of diverse stored structures}.
\newblock \emph{\bibinfo{journal}{Proceedings of the National Academy of
  Sciences}} \textbf{\bibinfo{volume}{112}}, \bibinfo{pages}{54--59}
  (\bibinfo{year}{2015}).
\newblock \urlprefix\url{https://doi.org/10.1073/pnas.1413941112}.

\bibitem{Pinto2024}
\bibinfo{author}{Pinto, D. E.~P.}, \bibinfo{author}{Ara\'ujo, N. A.~M.},
  \bibinfo{author}{\ifmmode~\check{S}\else \v{S}\fi{}ulc, P.} \&
  \bibinfo{author}{Russo, J.}
\newblock \bibinfo{title}{Inverse design of self-folding 3d shells}.
\newblock \emph{\bibinfo{journal}{Phys. Rev. Lett.}}
  \textbf{\bibinfo{volume}{132}}, \bibinfo{pages}{118201}
  (\bibinfo{year}{2024}).
\newblock \urlprefix\url{https://doi.org/10.1103/PhysRevLett.132.118201}.



\end{thebibliography}

\end{document}